\documentclass[prd,preprint,secnumarabic,showkeys,showpacs,amsmath,amssymb]{revtex4-1}

\usepackage{graphicx}
\usepackage{amsmath}
\usepackage{floatrow}
\begin{document}

\title{An FLRW Cosmology with a Chameleon Field}

\author{Amir Pouyan Khosravi Karchi}
 \email{amir.khosravi@mail.sbu.ac.ir}

\author{Hossein Shojaie}
 \email{h-shojaie@sbu.ac.ir}

\affiliation{Department of Physics, Shahid Beheshti University, G.C., Evin, Tehran 1983963113, Iran}

\begin{abstract}
In this manuscript, the field equations of a chameleon field in which the matter Lagrangian term is a general function of the scalar field as well as matter field, are derived. The equations are then expressed in Friedmann--Lema\^itre--Robertson--Walker~(FLRW) framework and the associated phase portraits and a power law solution are discussed in details. It is shown that why non-minimal coupling between the chameleon and matter fields leads to an energy transfer between the fields which consequently, affects the expansion rate of the universe. The transfer direction is determined by the second law of thermodynamics. The solution indicates that an accelerating expansion of the universe can be described as a result of the energy flow from the chameleon field to matter field.     
\end{abstract}



\maketitle

\section{Introduction}

The concordance model of the universe, or namely Lambda--Cold--Dark--Matter ($\Lambda$CDM) model, is in very good agreement with the data obtained from Planck space telescope~\cite{Planck1}; data which has provided a very high resolution image of Cosmic Microwave Background radiation~(CMB)~\cite{CMB1}. This data, in addition to a wide variety of cosmological observations such as Baryon Acoustic Oscillations~(BAO)~\cite{BAO}, Weak Lensing~\cite{WeakLensing}, Supernova type Ia~(SNIa)~\cite{SNIa,Riess}, 2dF Galaxy Redshift Survey~(2dFGRS)~\cite{2dFGRS} and DEEP2 Redshift Survey~\cite{DEEP2} at low and high redshifts respectively, has been used to tune $\Lambda$CDM model.                 
Accordingly, this model implies that the content of our $13.68$ billion year old universe is built up of $32\%$ non--relativistic matter and $68\%$ unknown isotropic and homogeneous distributed matter with equation of state $w=-1$, dubbed as dark energy~(DE)~\cite{LCDM}. Moreover, the universe turns out to be almost spatially flat, homogeneous and isotropic at its largest scale, and is recently experiencing an accelerating phase~\cite{LCDM-GEN}.
 
Although there are different models of DE~\cite{DE-MOD} and even different classifications of it concerning the equation of state~\cite{DE-CLASS}, other alternatives to general relativity (GR) are widely discussed in the literature in order to explain the nature of the late time cosmic acceleration. Among them, the scalar--tensor theories of gravitation, as extensions of Brans--Dicke~(BD) theory of gravitation, are the most well--known. These theories non--minimally couple a scalar field to the Einstein--Hilbert Lagrangian, to preserve the Mach principle. BD theory of gravitation mainly suffers to bring about a fine match with the observations in the solar system, within plausible value of its parameter~\cite{BD-SOL-SYS}. Other modifications of GR concern, for instance, the replacement of the scalar field with functions of other scalars engaged in the theory, such as $f(R)$~\cite{FR}, $f(T)$~\cite{FT} and $f(G)$~\cite{FG}, where $R$, 
$T$ and $G$ are the Ricci scalar, the torsion scalar and the Gauss--Bonnet term, respectively. These latter alternatives to GR are capable of explaining the inflation phase and the late time acceleration in a unified manner. 

The chameleon theory of gravitation~\cite{CHAM}, as another modification to GR, mainly differs from BD theory in that the scalar field is non--minimally coupled to matter field. In other words, the matter Lagrangian density depends on the scalar field as well as matter field and the metric. This allows the chameleon field to couple to the matter much stronger than gravity does and consequently making this kind of theory capable of being in agreement with observational constraints~\cite{CHAM-OBS}. 

In this manuscript, we investigate a chameleon theory which allows the scalar field to couple with both gravity and matter, which has been called the chameleonic generalized Brans-Dicke model (CGBD)~\cite{Faraj}, as an integration of both the chameleon and BD theories. In Section~2, rather than choosing a specified form of the matter Lagrangian term to explain the cosmic acceleration or other observational data, the most general form of Lagrangian~\cite{Khoury} is considered and the corresponding field equations are presented. Section~3 briefly provides the form of these equations in the Friedmann--Lema\^itre--Robertson--Walker~(FLRW) framework and also their associated phase portraits. In most of studied chameleon models, a specific choice of matter Lagrangian term enables one to easily solve the resulting field equations. This is the aim of Section~4 wherein a particular power law solution is provided. Finally, a conclusion is summarized in Section~5.

\section{General Field Equations}

Following the pioneer paper of Brans and Dicke~\cite{BD}, the BD Lagrangian is
\begin{equation}\label{lag1}
I=\int d^4x\sqrt{-g}(\phi R+L_m-\omega\frac{\phi_{,a}{\phi_,}^a}{\phi})\hspace{6pt},
\end{equation}
where $\omega$ is a positive constant. Dimensional analysis leads one to interpret
\begin{equation}
\phi\propto\frac{1}{ G_{\mathrm{eff}}}\hspace{6pt},
\end{equation}
in the context of standard cosmology. In addition, the matter Lagrangian density, is assumed to be only a function of matter field and the metric and independent of the field $\phi$. Varying the Lagrangian with respect to the metric and the scalar field, and combining the results, the corresponding field equations become:
\begin{equation}\label{rab1}
R_{ab}-\frac{1}{2}g_{ab}R=\frac{8\pi}{\phi}{T}_{ab}
+\frac{\omega}{\phi^2}\left(\phi_{,a}\phi_{,b}
-\frac{1}{2}g_{ab}\phi_{,c}{\phi_{,}}^c\right)+\frac{1}{\phi}\left(\phi_{,a;b}-g_{ab}{\phi_{,c;}}^c\right)\hspace{6pt},
\end{equation}

\begin{equation}\label{tab1}
{{T}_{ab;}}^b=0
\end{equation}
and
\begin{equation}\label{phaa1}
{\phi_{,a;}}^a=\frac{8\pi}{3+2\omega}T\hspace{6pt}.
\end{equation}

On the other hand, and as stated before, in a simple chameleon model, the Lagrangian is identical to the BD Lagrangian~(\ref{lag1}), except it is assumed that the term containing the matter Lagrangian density is also a function of $\phi$. Accordingly, one can consider the Lagrangian:
 \begin{equation}\label{lag}
I=\int d^4x\sqrt{-g}(\phi R+f(\phi)L_m-\omega\frac{\phi_{,a}{\phi_,}^a}{\phi})\hspace{6pt},
\end{equation}
which, for instance, has been studied as CGBD model in~\cite{Faraj} and~\cite{Faraj007}. Herein, we do not restrict the matter Lagrangian term to be separable for matter and the scalar field, as it is in~(\ref{lag}), and write down the Lagrangian as:  

\begin{equation}\label{lag2}
I=\int d^4x\sqrt{-g}(\phi R+L_m(\phi)-\omega\frac{\phi_{,a}{\phi_,}^a}{\phi})\hspace{6pt},
\end{equation}

where $L_m$ is a function of the field $\phi$ as well as matter field and the metric.
This assumption results in a modification of the field equations,~(\ref{rab1}),~(\ref{tab1}) and~(\ref{phaa1}), to the following equations respectively, namely,

\begin{equation}\label{rab2}
R_{ab}-\frac{1}{2}g_{ab}R=\frac{8\pi}{\phi}{T}_{ab}
+\frac{\omega}{\phi^2}\left(\phi_{,a}\phi_{,b}
-\frac{1}{2}g_{ab}\phi_{,c}{\phi_{,}}^c\right)+\frac{1}{\phi}\left(\phi_{,a;b}-g_{ab}{\phi_{,c;}}^c\right)\hspace{6pt},
\end{equation}

\begin{equation}\label{tab2}
{{T}_{ab;}}^b=\frac{1}{2}S\phi_{,a}
\end{equation}
and
\begin{equation}\label{phaa2}
{\phi_{,a;}}^a=\frac{8\pi}{3+2\omega}\left(T-\phi S\right)\hspace{6pt},
\end{equation}
where $S$ is defined as
\begin{equation}
S\equiv \frac{1}{\sqrt{-g}}\frac{\delta L_m}{\delta\phi}\hspace{6pt}.
\end{equation}

These equations are the general field equations of this theory. However, to have a specific solution, one has to apply the boundary conditions and/or symmetries to the field equations. In the next section the specific forms of these equations in the cosmological framework are derived.

\section{Chameleon Field In FLRW Framework}

\subsection{Field Equations}

The cosmological principle states that, in the universe when viewed on sufficiently large scales, there is no preferred place or direction in space. In other words, the universe can be approximated as homogeneous and isotropic in large scales. This principle is in accordance with the large scale observational evidences and can be applied to all alternatives to GR. Following the cosmological principle, the FLRW metric, namely
\begin{equation}
ds^2=-dt^2+a^2(t)
\left(\frac{dr^2}{1-kr^2}+r^2d\Omega^2\right) ,
\end{equation} 
is the metric best applicable to these scales in the universe, for in this frame all variables can be regarded as a function of cosmic time only. The observational evidences~\cite{FLATNESS} also convince one to assume spatial flatness of the universe in this frame. In addition, the matter field is considered to be perfect fluid with the equation of state
\begin{equation}
p=(\gamma-1)\rho\hspace{6pt}.
\end{equation}

The field equations~(\ref{rab2}) to~(\ref{phaa2}) in this framework, after some mathematical manipulations, are simplified to
\begin{equation}
\left(\frac{\dot a}{a}\right)^2=\frac{8\pi}{3\phi}\rho+\frac{\omega}{6}\left(\frac{\dot\phi}{\phi}\right)^2-\frac{\dot a}{a}\frac{\dot\phi}{\phi}\label{90}\hspace{6pt},
\end{equation}
\begin{equation}
\frac{\ddot a}{a}=-\frac{8\pi}{3\phi}\left[\rho\left(\frac{3+\omega}{3+2\omega}\right)+3p\left(\frac{\omega}{3+2\omega}\right)\right]-\frac{\omega}{3}\left(\frac{\dot\phi}{\phi}\right)^2+\frac{\dot a}{a}\frac{\dot\phi}{\phi}+\frac{4\pi}{3+2\omega}S\label{91}\hspace{6pt},
\end{equation}
\begin{equation}
\frac{\ddot\phi}{\phi}+3\frac{\dot a}{a}\frac{\dot\phi}{\phi}=\frac{8\pi}{(3+2\omega)\phi}(\rho-3p)-\frac{8\pi}{3+2\omega}S\label{101}
\end{equation}
and
\begin{equation}
\dot\rho+3\gamma\frac{\dot a}{a}\rho=\frac{1}{2}S\dot\phi\label{102}\hspace{6pt},
\end{equation}
where over--dot denotes derivation with respect to cosmic time $t$.

Defining
\begin{equation}
\rho_\phi\equiv\frac{\omega\dot\phi^2}{16\pi\phi}\label{100}
\end{equation}
lets one rewrite equation~(\ref{phaa2}) in the following form  
\begin{equation}
\dot\rho_\phi+6\frac{\dot a}{a}\rho_\phi=-\frac{1}{16\pi}R\dot\phi-\frac{1}{2}S\dot\phi\label{109}\hspace{6pt},
\end{equation}
which is similar to the Brans--Dicke scalar field equation except for the second term on the right hand--side. 

Equations~(\ref{102}) and~(\ref{109}) show that the densities $\rho$ and $\rho_{\phi}$ are not conserved. In other words, if one considers the general form of the continuity equation for an hypothetical quantity density $\rho$, that is
\begin{equation}
\dot{\rho}+3\gamma\frac{\dot{a}}{a} \rho =\Sigma
\label{con}
\end{equation}
then $\Sigma=0$ denotes a conserved quantity, and $\Sigma>0$ ($\Sigma<0$) corresponds to creation(annihilation) of the quantity related to the $\rho$. 

A comparison between equations~(\ref{102}) and~(\ref{con}) shows that the term $\frac{S\dot{\phi}}{2}$ is behaving like $\Sigma$ in equation~(\ref{con}), and hence, is playing the role the source or sink for the energy density of the scalar field which results in an energy transfer between the scalar and matter fields. One can define the creation era as an epoch in which the right hand--side of the equation~(\ref{102}) is positive. In other words, in this era the energy is transferred from the scalar field to the matter field. Conversely, we label the epoch in which the energy is transferred in opposite direction, that is from matter field to the scalar field, as the annihilation era. Equation~(\ref{109}), in addition, shows that the scalar field behaves just like a fluid with the equation of state $p=\rho$.

\subsection{Qualitative Analysis}

To study the dynamical behaviour of the universe according to the equations above,  one can introduce the following dynamical variables
\begin{equation}
X\equiv\frac{\dot a}{a}\hspace{6pt},\hspace{6pt}Y\equiv\frac{\dot\phi}{\phi}\hspace{6pt}\text{and}\hspace{6pt}Z\equiv\frac{\rho}{\phi}\hspace{6pt}.\label{200}
\end{equation}
In terms of these new variables, equations~(\ref{90}),~(\ref{91}) and~(\ref{101}) become
\begin{equation}
X^2=\alpha Z+\frac{\omega}{6}Y^2-X Y\hspace{6pt},\label{201}
\end{equation}
\begin{equation}
\dot X-X^2=-\alpha AZ-\frac{\omega}{3} Y^2+X Y+BS\label{202}
\end{equation}
and
\begin{equation}
\dot Y-Y^2+3X Y=2BCZ- 2BS\hspace{6pt},\label{203}
\end{equation}
where 
\begin{equation}
\alpha=\frac{8\pi}{3}\hspace{6pt},\hspace{6pt}
A=\frac{3-2\omega+3\gamma}{3+2\omega}\hspace{6pt},\hspace{6pt}
B=\frac{4\pi}{3+2\omega}\hspace{6pt}\text{and}\hspace{6pt}
C=2-3\gamma\hspace{6pt}.\label{280}
\end{equation}

Now, by defining
\begin{eqnarray}
Q_1\equiv\frac{B}{\alpha}\left(1-2\frac{X}{Y}\right)\hspace{6pt},\hspace{6pt}\\
Q_2\equiv1+\frac{2B\omega}{3\alpha}-\frac{2B}{\alpha}\frac{X}{Y}\hspace{6pt},\hspace{6pt}\\
Q_3\equiv1-\frac{B}{\alpha}-\frac{2B}{\alpha}\frac{X}{Y}\hspace{6pt}
\end{eqnarray}
and
\begin{eqnarray}
Q_4\equiv\frac{B}{\alpha}\left(\frac{X}{Y}-\frac{\omega}{3}\right)\hspace{6pt},
\end{eqnarray} 
and eliminating $Z$ between relations~(\ref{201}),~(\ref{202}) and~(\ref{203}), one gets

\begin{eqnarray}
\dot X&=&\left(Q_3-\frac{2Q_1Q_4}{Q_2}\right)^{-1}\left[\left(\frac{A\omega}{6}-\frac{\omega}{3}+\frac{Q_4}{Q_2}-\frac{B\omega}{6\alpha^2}-\frac{B Q_4\omega}{6 \alpha^2Q_2}-\frac{\gamma\omega}{2\alpha}\frac{X}{Y}\right)Y^2\right.\nonumber\\
&+&\left(1-A+\frac{B}{\alpha^2}+\frac{3\gamma}{2\alpha}\frac{X}{Y}\right)X^2\nonumber\\
&+&\left.\left(1-A-\frac{3Q_4}{Q_2}+\frac{B}{\alpha^2}+\frac{B Q_4}{Q_1\alpha^2}+\frac{3\gamma}{\alpha}\frac{X}{Y}\right)XY\right]\label{300}
\end{eqnarray}

and

\begin{eqnarray}
\dot Y&=&\left(1-2\frac{Q_1}{Q_2Q_3}\right)^{-1}\left\lbrace\left[\frac{1}{Q_2}\left(1-\frac{BC}{3}-\frac{B\omega}{6\alpha^2}-\frac{3B\omega}{6\alpha^2}\frac{X}{Y}\right)+\frac{Q_1\omega}{Q_2Q_3}\left(\frac{A}{3}-\frac{2}{3}-\frac{B}{3\alpha^2}-\frac{B\gamma}{2\alpha}\frac{X}{Y}\right)\right]Y^2\right.\nonumber\\
&+&\left[\frac{1}{Q_2}\left(\frac{2BC}{\alpha}+\frac{B}{\alpha^2}+\frac{3B}{2\alpha^2}\frac{X}{Y}\right)+\frac{Q_1}{Q_2Q_3}\left(2-2A+\frac{2B}{\alpha^2}+\frac{3B(\gamma+1)}{\alpha^2}\frac{X}{Y}\right)
\right]X^2\nonumber\\
&+&\left.\left[(\frac{1}{Q_2}\left(-3+\frac{2BC}{\alpha}+\frac{B}{\alpha^2}+\frac{3B}{2\alpha^2}\frac{X}{Y}\right)+\frac{Q_1}{Q_2 Q_3}\left(-2A+\frac{2B}{\alpha^2}+\frac{3B\gamma}{\alpha^2}\frac{X}{Y}\right)\right]XY\right\rbrace\label{301}.            
\end{eqnarray}

\begin{figure}[h!]
\includegraphics[scale=0.6]{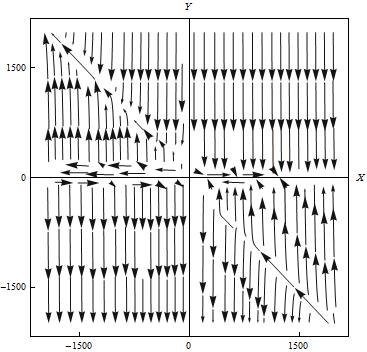}
\includegraphics[scale=0.5]{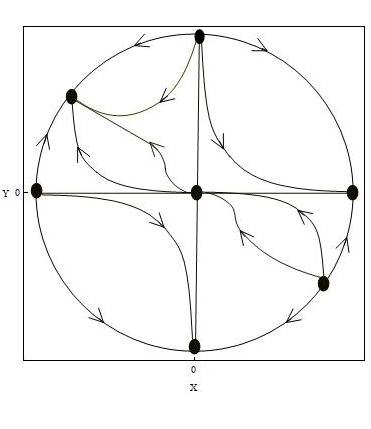}
\caption{The phase portrait for $\omega=-1.49$ and $\gamma=1$.This case shows the evolution of a cold dust dominant universe and includes both the inflationary phase and late time acceleration of the universe for different initial conditions}
\label{fig:fig12}
\end{figure}

\begin{figure}[h!]
\includegraphics[scale=0.6]{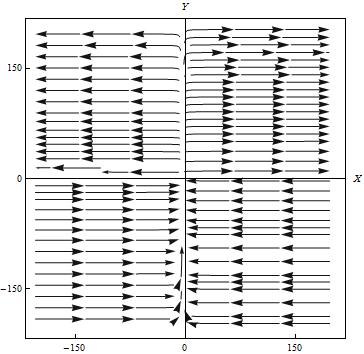}
\includegraphics[scale=0.548]{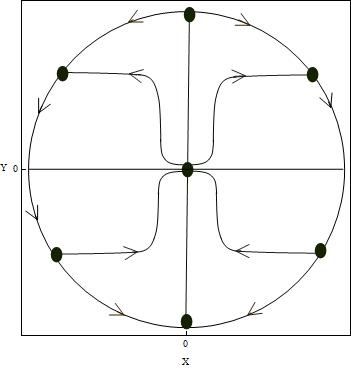}
\caption{The phase portrait for $\omega =-1.49$ and $\gamma=0$.This case shows a dark matter dominant universe and includes the curves which are manifestation of the late time acceleration}
\label{fig:fig13}
\end{figure}

\begin{figure}[h!]
\includegraphics[scale=0.6]{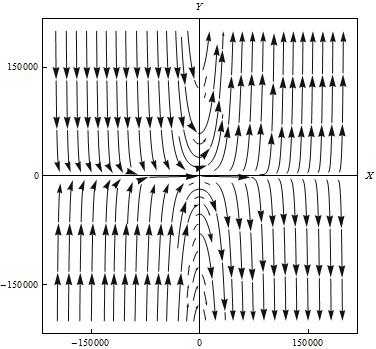}
\includegraphics[scale=0.548]{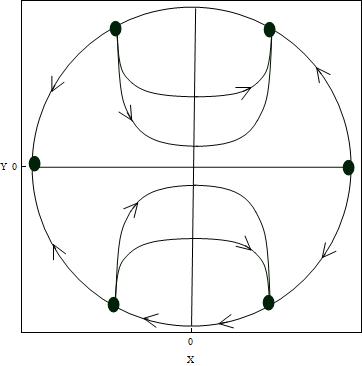}
\caption{The phase portrait for $\omega=50000$ and $\gamma=1$.The phase diagrams show the evolution of a cold dark matter dominated universe and they demonstrate signs of  a late time acceleration phase.}
\label{fig:fig14}
\end{figure}

These two equations can be used to analyze the universe as a dynamical system. However, since the dependence of the matter Lagrangian term on the scalar field is not specified, one can only have a qualitative analysis of the equations by means of $\gamma$ and $\omega$. It should be mentioned that one cannot expect this qualitative description to produce exactly the observed cosmological eras of the universe, since in this analysis, some assumptions, such as the constancy of the equation of state of the matter field during the evolution of the universe, have been used which do not manifest the reality. Regarding the definitions of the dynamical parameters, relations~(\ref{200}), $X$ is the Hubble parameter and so, any exponential expansion of the universe, such as the inflation phase and the late time acceleration, is represented by vertical sections of the integral curves. The big bang, on the other hand, corresponds to $X\rightarrow\infty$. As examples, the phase portraits of three cases ($\omega=-1.49$,$\gamma=1$),($\omega=-1.49$,$\gamma=0$) and ($\omega=50000$,$\gamma=1$) are provided in Figures~\ref{fig:fig12},~\ref{fig:fig13} and~\ref{fig:fig14}, respectively. The value of $\omega=-\frac{3}{2}$ and $\omega=50000$ correspond to a conformal invariant BD theory~\cite{CONINV} and GR limit respectively. 

\vspace{600pt}

\section{A Typical Solution}

To have a typical solution for the model in FLRW framework, let us consider the power law evolution of the scale factor and the scalar field as an example, that is
\begin{equation}\label{assumptions}
\phi=\phi_0 t^m
\end{equation}
and
\begin{equation}
 a=a_0 t^n,
\end{equation}
and checkout the consequences dictated by our field equations. Consequently, equations~(\ref{91}) to~(\ref{102}) give:
\begin{equation}\label{rhos}
\rho=\rho_0 t^{m-2}  
\end{equation}
and
\begin{equation}\label{rhos1}
S=S_0 t^{-2}.
\end{equation}
Rewriting equation~(\ref{rhos}) as:
\begin{equation}
\rho=\frac{\rho_0}{ t^2}t^{m}  
\end{equation}
and taking into account that $\frac{\rho_0}{ t^2}$ is the matter density in standard cosmology, it becomes clear that $m>0$ and $m<0$ denote the creation and annihilation eras, respectively. It is worth noting that the parameters $a_0$, $\phi_0$, $\rho_0$ and $S_0$ can be considered as the values of the corresponding quantities at the present time. 
 
By substituting the above relations into the field equations ~(\ref{90}) to~(\ref{102}), one gets
\begin{equation}
n^2=\frac{8\pi}{3} \frac{\rho_0}{\phi_0}+\frac{\omega m^2}{6}-nm ,
\end{equation}
\begin{equation}
n(n-1)=\frac{-8\pi}{3} \frac{\rho_0}{\phi_0}[\frac{3+\omega}{3+2\omega}+\frac{3(\gamma-1)\omega \rho_0}{(3+2\omega)\phi_0}]-\frac{\omega m^2}{3}+nm+\frac{4\pi S_0}{3+2\omega} ,
\end{equation}
\begin{equation}
m(m-1)+3nm=\frac{8\pi}{3+2\omega} [\frac{\rho_0}{\phi_0}-\frac{3(\gamma-1)\omega \rho_0}{\phi_0}]-\frac{8\pi S_0}{3+2\omega} 
\end{equation}
and
\begin{equation}\label{exchange}
(m-2) \rho_0+3\gamma \rho_0 n=\frac{S_0 \phi_0 m}{2}.
\end{equation}
The latter equation can be rearranged as
\begin{equation}\label{line}
n=Dm+\frac{2}{3 \gamma},
\end{equation}
where $D =\frac{1}{\gamma}(\frac{s_0 \phi_0}{6  \rho_0}-\frac{1}{3 })$. 

It's worth noting that, if one additionally assumes the scalar field $\rho_{\phi}$, as well as the matter field, to obeys the weak energy condition, then according to relation~(\ref{100}), $\phi_0$ will be non--negative. Moreover, as discussed in lines after equation~(\ref{rhos}), the energy transfer direction is solely determined by the sign of $m$. Hence, putting these all together, the right hand--side of equation~(\ref{exchange}) implies that $S_0$ should be positive regardless of sign of $m$. As a summary, equations~(\ref{102}) and~(\ref{109}) show that the term $\frac{1}{2}S \dot{\phi}$, which is already introduced as a rate of energy flow between the scalar and matter fields, transfers the energy from the former to the latter for, $m>0$, and in the opposite direction for, $m<0$.

The condition for the expansion and an accelerating expansion of the universe can be obtained by imposing required constraints on $n$ or $m$ in equation~(\ref{line}). However, taking into account the second law of thermodynamics~\cite{Bisabr}, that is by non--decreasing behavior of the entropy of matter field, one can conclude $m\geq0$ for an expanding universe.

\begin{figure}[H]
\includegraphics[scale=0.47]{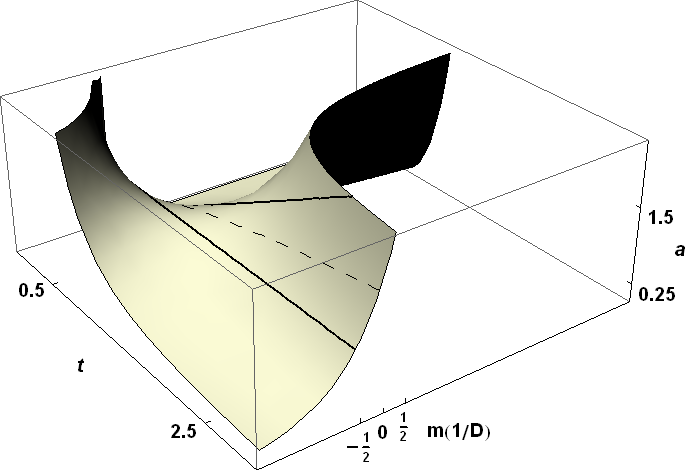}
\caption{Scale factor versus time and $m$ for a radiation dominated universe with $D>0$. The dashed line, $m=0$, indicates the borderline between the matter creation and annihilation regions. The black lines separate different states of the evolution of the universe.}
\label{fig:1}
\end{figure}

\begin{figure}[H]
\includegraphics[scale=0.47]{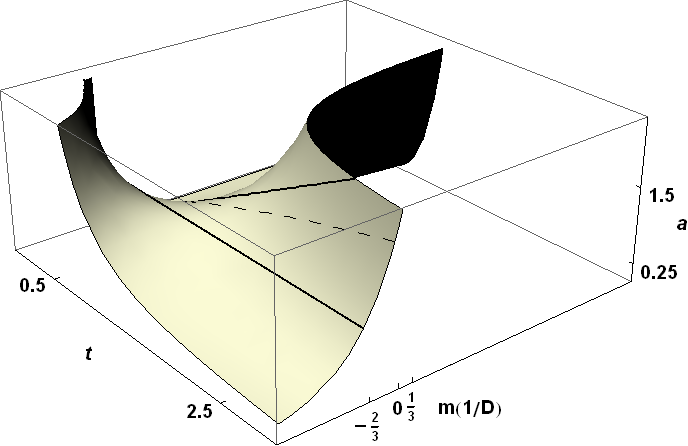}
\caption{Scale factor versus time and $m$ for a matter dominated universe with $D>0$. The dashed line, $m=0$, indicates the borderline between the matter creation and annihilation regions. The black lines separate different states of the evolution of the universe.}
\label{fig:2}
\end{figure}

\begin{figure}[H]
\includegraphics[scale=0.47]{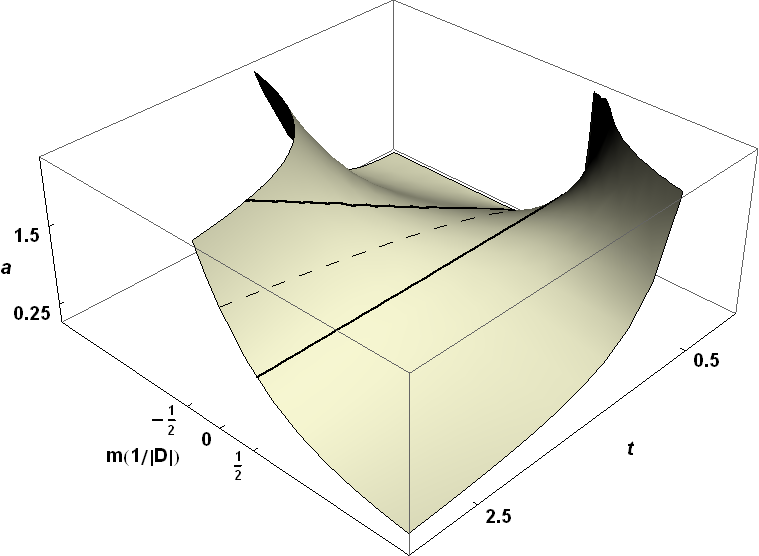}
\caption{Scale factor versus time and $m$ for a radiation dominated universe with $D<0$. The dashed line, $m=0$, indicates the borderline between the matter creation and annihilation regions. The black lines separate different states of the evolution of the universe.}
\label{fig:3}
\end{figure}

\begin{figure}[H]
\includegraphics[scale=0.47]{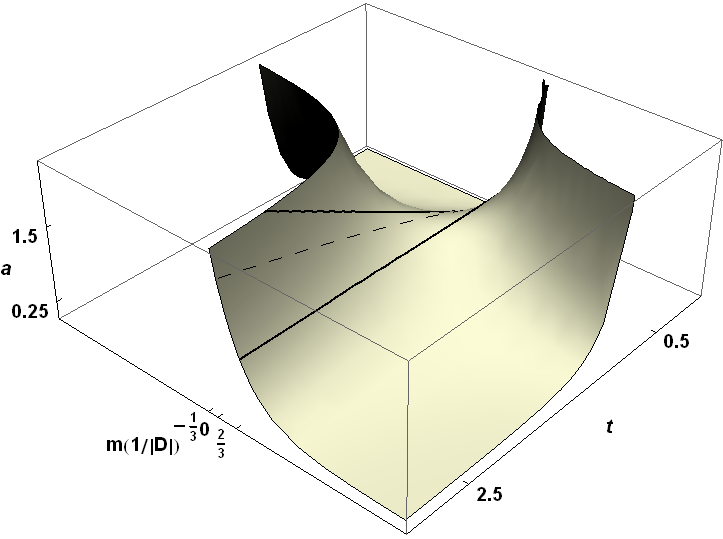}
\caption{Scale factor versus time and $m$ for a matter dominated universe with $D<0$. The dashed line, $m=0$, indicates the borderline between the matter creation and annihilation regions. The black lines separate different states of the evolution of the universe.}
\label{fig:4}
\end{figure}

Tables~\ref{tab:1} to~\ref{tab:4} and associated Figures~\ref{fig:1} to~\ref{fig:4} summarize different possible eras in the evolution of the universe with respect to intervals of $m$, separately for radiation and matter dominated epochs. 

\begin{table}[h]
\caption {Different phases of a radiation dominated universe with respect to the intervals of $m$ for  $D>0$} \label{tab:1} 
\begin{tabular}{|c| c |c|}
\hline
$m(1/D)$      & Scale factor evolution & Matter creation rate \\  \hline  \hline
$m\textless-1/2$  & Shrinking & $-$ \\ \hline 
$-1/2\textless m\textless0$  & Expanding  & $-$ \\ \hline 
$m=0$  & Expanding  & $0$ \\ \hline  
$0\textless m\textless1/2$  & Expanding   & $+$ \\ \hline 
$1/2\textless m$     &  Accelerating Expansion  & $+$ \\ \hline 
\end{tabular}
\end{table}

\begin{table}[h]
\caption {Different phases of a matter dominated universe with respect to the intervals of $m$ for the case $D>0$} \label{tab:2} 
\begin{tabular}{ |c| c| c|}
\hline
$m(1/D)$                 & Scale factor evolution & Matter creation rate \\ \hline \hline
$m\textless-2/3$       & Shrinking         &   $-$   \\ \hline
$-2/3\textless m\textless0$        & Expanding  & $-$  \\ \hline
$m=0$        & Expanding  & $0$  \\ \hline
$0\textless m\textless1/3$        & Expanding  & $+$  \\ \hline
$1/3\textless m$               & Accelerating Expansion   &  $+$  \\ \hline
\end{tabular}
\end{table}

\begin{table}[h]
\caption {Different phases of a radiation dominated universe with respect to the intervals of $m$ for the case $D<0$} \label{tab:3}
\begin{tabular}{|c| c| c|}
\hline
$m(1/|D|)$                 & Scale factor evolution & Matter creation rate \\ \hline \hline
$1/2\textless m$  & Shrinking         &        $+$   \\ \hline
$0\textless m \textless1/2$ & Expanding         &         $+$        \\ \hline
$m=0$ & Expanding        &        $0$        \\ \hline
$-1/2\textless m \textless0$ & Expanding         &      $-$       \\ \hline
$m \textless -1/2$              & Accelerating Expansion  &  $-$       \\ \hline
\end{tabular}
\end{table}

\begin{table}[h]
\caption {Different phases of a matter dominated universe with respect to the intervals of $m$ for the case $D<0$} \label{tab:4}
\begin{tabular}{|c|c|c|}
\hline
$m(1/|D|)$                  & Scale factor evolution & Matter creation rate  \\ \hline \hline
$2/3\textless m$     & Shrinking         &       $+$      \\ \hline
$0\textless m \textless2/3$      & Accelerating    &    $+$    \\ \hline
$m=0$      & Accelerating    & $0$      \\ \hline
$-1/3\textless m \textless0$      & Accelerating    &  $-$     \\ \hline
$m \textless -1/3$        & Accelerating Expansion    &  $-$   \\ \hline
\end{tabular}
\end{table}

\section{Conclusion}
In this manuscript, a CGBD theory, in which a scalar field is non--minimally coupled to both matter and gravity, has been studied and the corresponding field equations in the most general form have been derived. A particular solution has been then discussed and consequences of the non--minimal coupling on the expansion rate of the universe and the matter creation rate are provided. To be more precise, an accelerating expansion of the universe in a matter dominated era, that may be put in agreement with the concordance model and second law of thermodynamics, has been represented by the interval $m>\frac{1}{3}$, which can be interpreted as a consequence of the transfer of energy from the scalar field to matter field. Further studies can include different types of solutions for equations such as exponential solution, different model parameters with different initial conditions and also a potential term added to the Lagrangian.

\section*{Acknowledgment}

HS would like to appreciate the research council of Shahid Beheshti University for the financial support.

\setlength{\belowdisplayskip}{1cm}
\end{document}